\documentclass[12pt]{article}
\usepackage[authoryear, round]{natbib}
\usepackage{amsmath}
\usepackage{graphicx,psfrag,epsf}
\usepackage{enumerate}

\newcommand{\blind}{0}
\usepackage{amsfonts}
\usepackage{bm}

\usepackage{graphicx} 
\usepackage{subfig,tikz}
\usepackage{tkz-graph} 

\usetikzlibrary{arrows.meta} 
\usetikzlibrary{positioning}
\usetikzlibrary{arrows,shapes.arrows,shapes.geometric,shapes.multipart,decorations.pathmorphing, positioning,shapes.swigs, calc}
\tikzset{edge/.style = {->,> = latex'}}

\newcommand{\indep}{\perp\hskip -7pt \perp }

\usepackage[hidelinks, breaklinks]{hyperref}

\usepackage{cleveref}

\begin{document}

\def\spacingset#1{\renewcommand{\baselinestretch}%
{#1}\small\normalsize} \spacingset{2}


\if0\blind
{
  \title{\bf Complementary strengths of the Neyman-Rubin and graphical causal frameworks}
  \author{Tetiana Gorbach\thanks{
    This work was supported by the \textit{Marianne and Marcus Wallenberg Foundation under Grant number MMW 2021.0020, the Swedish Research Council under grant number 2016-00703, and by the Research Council of Finland under grant number 368935.}}\hspace{.2cm}\\
    Department of Statistics, USBE, Ume{\aa} University\\
    and \\
    Xavier de Luna \\
    Department of Statistics, USBE, Ume{\aa} University\\
    and\\
    Juha Karvanen \\
    Department of Mathematics and Statistics, University of Jyvaskyla\\
    and\\
    Ingeborg Waernbaum\\
    Department of Statistics, Uppsala University}
    \date{}
  \maketitle
} \fi

\if1\blind
{
  \bigskip
  \bigskip
  \bigskip
  \begin{center}
    {\LARGE\bf Title}
\end{center}
  \medskip
} \fi

\bigskip
\newpage
\begin{abstract}
This article contributes to the discussion on the relationship between the Neyman-Rubin and the graphical frameworks for causal inference. 
We present specific examples of data-generating mechanisms — such as those involving undirected or deterministic relationships and cycles — where analyses using a directed acyclic graph are challenging, but where the tools from the Neyman-Rubin causal framework are readily applicable.  
We also provide examples of data-generating mechanisms with M-bias, trapdoor variables, and complex front-door structures, where the application of the Neyman-Rubin approach is complicated, but the graphical approach is directly usable. 
The examples offer insights into commonly used causal inference frameworks and aim to improve comprehension of the languages for causal reasoning among a broad audience.
\end{abstract}

\noindent%
\vfill

\newpage
\spacingset{2} 
\section{Introduction}

We consider two prevalent frameworks for causal reasoning and inference: the Neyman-Rubin causal model, also known as the potential outcome framework \citep{neyman1923applications, rubin1974estimating}, and the graphical causal framework \citep{pearl1995causal, pearl2009causality}.  Both frameworks provide structured approaches to causal inference.

The Neyman-Rubin causal framework defines causal effects through comparisons between potential outcomes - the outcomes that would be observed for a unit under each possible treatment value.
 For the identification and estimation of causal effects, this framework relies on several key assumptions: consistency, that connects the potential and the observed outcomes \citep{cole2009consistency},  the stable unit treatment value assumption (SUTVA, \citealp{rubin1980randomization}), positivity (all units having non-zero probability of being treated), and unconfoundedness (no unmeasured confounding of the treatment-outcome relationship, also referred to as the ignorability of the treatment assignment,   \citealp{rosenbaum1983central}).  

The graphical causal framework, on the other hand, uses graphs to represent the causal structures and associated do-calculus to identify causal effects \citep{pearl2009causality}. The predominantly used graph and the core of the graphical framework is a directed acyclic graph (DAG).   
Each node in a causal DAG represents a variable in a causal system, and each edge  denotes a causal relationship between variables \citep[p.12]{pearl2009causality}. 
Graphical framework relies on the assumptions that connect  probability distributions and conditional independencies implied by a graph \citep{dawid2010beware}, such as  Markov condition \citep{pearl1995theory} and faithfulness \citep{spirtes1993causation}.

The relationship between the frameworks' philosophy has been extensively discussed in the scientific community.
\citet{pearl_potential_outcomes}, for example, asserts that the two are ``logically equivalent" and can be used ``interchangeably and symbiotically". \citet{imbens2020potential} suggests that ``these frameworks are complementary, with different strengths that make them particularly appropriate for different questions". 
 \citet{ibeling2024comparing} conclude that ``$\ldots$ the two frameworks are productively compatible, while also suggesting distinctive perspectives on problems of causal inference." \cite{markus2021causal}, on the other hand, argues for at most weak but no strong equivalence between the frameworks (see also a commentary of \citealp{weinberger2023comparing}).

In practice,  the choice between frameworks in a specific application may be informed by the amount of qualitative information about the system, philosophical traditions within the research field, and the type of research question.
For example, the graphical framework requires specification of the conditional independencies within structural equations in a stable system.  
The Neyman-Rubin framework, on the other hand, is concerned with the causal mechanisms of the treatment and potential outcomes, typically by identifying common causes to satisfy the unconfoundedness assumption. 
This approach may require fewer assumptions as it does not require specifying all conditional independencies in the considered data-generating process, as in the graphical framework.
For example, in randomized controlled trials, where the unconfoundedness is fulfilled by design, the potential framework is a natural choice.   
In contrast, specifying a graph may enhance the understanding of complex structures and help identify causal effects when this is challenging in the Neyman-Rubin framework. 

Other frameworks for causal inference, which are not discussed in this paper, include models based on Single-World Intervention Graphs (SWIGs, \citealp{richardson2013single}), which are based on DAGs and incorporate counterfactuals into graphs; models with axiomatization grounded in measure theory \citep{park2023measure}; a decision-theoretic approach to causal inference \citep{dawid2021decision}; Lewis' Theory of Counterfactuals \citep{lewis1973causation, lewis1973counterfactuals}; and the framework of \citet{heckman2015causal} based on Haavelmo's work \citep{haavelmo1943statistical, haavelmo1944probability}.

In this article, we expand on the distinctive perspectives of the approaches by providing straightforward examples of how the graphical and the Neyman-Rubin frameworks can complement each other.  While we do not aim to assert the mathematical equivalence of the frameworks, we focus on their practical use in specific cases.  We present three data-generating mechanisms that cannot be represented as-is by a DAG and require more complex graphical models, while the Neyman-Rubin causal framework offers more straightforward tools. We also present real-world situations where the structures from the examples might occur, along with possible solutions in the graphical causal framework. We then provide examples of data-generating mechanisms that can readily be handled by using DAGs but are challenging for the Neyman-Rubin framework. The examples are implemented in R \citep{r2022} and the corresponding code is available in the online supplementary material. 
The examples and discussion are provided from our background of working both as theoretical and applied statisticians in fields of epidemiology, forestry, social sciences, and psychology, thus expanding the complementary views from the economics’ perspective provided by Imbens (2020).
Our goal is to foster understanding of the frameworks among a wider audience through a neutral and illustrative discussion.


\section{Examples where the Neyman-Rubin causal model provides simpler tools} \label{section:po_easier_than_dags}

In the Neyman-Rubin causal framework,  covariate adjustment allows for the estimation of the causal effects regardless of the nature of the relationship between the pre-treatment covariates as long as the consistency,  SUTVA, positivity, and unconfoundedness assumptions are fulfilled.

In contrast, the graphical causal framework requires that relationships between variables be both directed and acyclic to represent data-generating mechanisms using a DAG, as the name implies. 
Also, a DAG typically represents a data-generating mechanism where endogenous variables are functions of random disturbance terms and other variables in the DAG, and not just deterministic functions of other variables in this DAG \citep[Definition 2.2.2]{pearl2009causality}.

In this section, we give examples of data-generating mechanisms that exhibit undirected, cyclic, and deterministic relationships that are challenging to represent with a DAG. Consequently,  advanced methods that use other graphs than DAGs are required for their analyses within the graphical framework, while the Neyman-Rubin causal model offers a simpler alternative.


\subsection*{Example 1. Undirected relationships}

Non-causal associations can, for instance, occur in processes used to characterize marked spatial point patterns \citep{diggle2013statistical}. For example, in a forestry application, we might want to investigate the effect of partially harvesting some trees on the future growth of the forest. The volume of a tree can be associated with the volumes of its neighboring trees through non-causal relationships due to, for example, influence zone overlap \citep{sarkka2006analysis}, while showing no such association with trees located farther away.



To illustrate this, take the following (simplified) numerical example involving an i.i.d. sample of a binary treatment $A,$ pre-treatment covariates $C_1, \ldots, C_4$, an observed outcome $Y, $ and the corresponding potential outcomes $Y(0)$ and $Y(1),$ where
\begin{align} \label{eq:model_treatment_outcomes} 
    \begin{cases}
        A \sim Bern \left(\frac{\exp(C_1 + C_2)}{1 + \exp(C_1 + C_2)}\right)\\
        Y(0) = 4 + C_1 + C_2 + \varepsilon_0\\
        Y(1) = 2 + C_1 + C_2 + \varepsilon_1\\
        Y = AY(1) + (1-A)Y(0), 
    \end{cases}
\end{align}
and
$(\varepsilon_0, \varepsilon_1)' \sim \mathcal{N} \left(0, I_2 \right)$ independent of treatment $A $ and covariates $C_1, \ldots, C_4$, and $I_2$ is the $2\times 2$ identity matrix. 
The undirected relationships between the pre-treatment covariates $C_1, C_2, C_3,$ and $C_4$ are expressed by a multivariate normal distribution:
\begin{align*} 
    \begin{pmatrix}
        C_1\\
        C_2\\
        C_3\\
        C_4
    \end{pmatrix}  \sim \mathcal{N} 
    \left(
        \begin{pmatrix}
            0\\
            0\\
            0\\
            0
        \end{pmatrix}, 
        \begin{pmatrix}
            1 & \rho & \rho & a_\rho \\
            \rho & 1 & a_\rho & \rho \\
            \rho & a_\rho  & 1 & \rho \\
            a_\rho & \rho & \rho &1 
        \end{pmatrix}
    \right), 
\end{align*}
where $0 < \rho < 1$ and $a_\rho = (\sqrt{8 \rho^2 + 1} - 1)/2.$
For this data-generating mechanism, the consistency assumption within the Neyman-Rubin causal framework is fulfilled by construction since $Y = Y(1)$ if $A=1$ and $Y = Y(0)$ if $A=0.$
Theoretically, the positivity assumption is fulfilled since the probability of receiving treatment level $A=1$ is positive and less than 1 for any values of $C_1$ and $C_2, $  since $0<\frac{\exp(C_1 + C_2)}{1 + \exp(C_1 + C_2)}<1.$ 
Note, however, that high (and low) values of $C_1+C_2$ might complicate the estimation from finite samples since the observed proportions of units receiving treatment level $A=1$ can be 1 or 0.  
The SUTVA assumption is fulfilled since the sample is i.i.d, and thus the potential outcomes for a unit are independent of the treatment assignments for other units. 
The unconfoundedness assumption is satisfied because $(Y(0), Y(1)) \indep A |C_1, C_2, C_3, C_4, $  as $(\varepsilon_0, \varepsilon_1) $ are independent of both treatment $A$ and the covariates $C_1, \ldots, C_4$.  
As a result, the average causal effect can be identified by adjusting the observed outcome $Y$ for confounders $C_1, \ldots, C_4$ \citep{rubin1974estimating}.

This example is difficult to address using a DAG, as the underlying data-generating mechanism cannot be captured by a traditional DAG structure \textit{as it is} since the relationships between $C_1, C_2, C_3,$ and $C_4$ are undirected (see \Cref{fig: example_undirected} for an illustration). In fact, no DAG can produce the relations between $C_1, \ldots, C_4$ in Example 1 \citep[p.83]{koller2009probabilistic}.

In certain situations, associative relationships can be analyzed within the graphical framework. For example, clustering variables could sometimes transform the graph with undirected relationships into a DAG. When the undirected relationship between some variables is irrelevant for identifying the causal effect of interest, these variables might be represented as one node in a DAG/SWIG \citep{anand2023causal, tikka2023clustering}. 
For example, the data-generating mechanism in the above-mentioned example can be represented by a DAG with all confounders $C_1, \ldots, C_4$  represented as one node $\bf{C}$ as in \Cref{fig: example_cluster}. It is easy to check that the traditional back-door criterion is fulfilled for the graph with one node $\bf{C}.$ Consequently, the average causal effect of treatment $A$ on the outcome $Y$, for example, is identified by adjusting $Y$ for $C_1, \ldots, C_4. $ However,  a new clustering concept is needed because the clustering methodology by \citet{tikka2023clustering} is developed only for directed graphs. 

When the association between observed variables is due to an unobserved confounder, more advanced graphical models compatible with acyclic-directed mixed graphs can be utilized (see, for example,  \citealp{bhattacharya2022semiparametric}). 
Also, \citet{maathuis2015generalized} and \citet{perkovic2015complete} generalized the traditional back-door criterion to completed partially directed acyclic graphs, classes of DAGs with non-directed edges representing the existence in a class of at least two DAGs: one with an arrow in one direction and another with the arrow in the opposite direction \citep{andersson1997characterization, meek2013causal, spirtes1993lecture}. This generalization can also be applied when associations between the observed variables are due to an unobserved variable(s), as in maximal ancestral graphs 
\citep{richardson2002ancestral, richardson2003causal}, or partial ancestral graphs representing classes of maximal ancestral graphs with the same conditional independences between the observed variables \citep{ali2012towards, richardson2002ancestral}.

\subsection*{Example 2. Cyclic relationships} 
The cyclic relationship between some of the variables does not restrict analyses in the Neyman-Rubin framework as long as the assumptions required for causal effect identification and estimation are fulfilled.

Consider the data-generating mechanism where equation~\eqref{eq:model_treatment_outcomes} holds for $A$ and $Y$, and that includes a cycle between the pre-treatment covariates:
\begin{align*}
    \begin{cases}
        &(\varepsilon_0, \varepsilon_1, \varepsilon_2, \varepsilon_3)' \sim \mathcal{N} \left(0, I_4\right)\\
        & C_1 = 0.1 C_2 + \varepsilon_2\\
        & C_2 = 0.1 C_1 + \varepsilon_3,
    \end{cases}
\end{align*}
where $I_4$ is a $4\times 4$ identity matrix.
As in Example 1, the consistency, positivity, and SUTVA assumptions are fulfilled. The potential outcomes are independent of the treatment given covariates $C_1$ and $C_2,$ $(Y(0), Y(1)) \indep A |C_1, C_2,$ since $(\varepsilon_0, \varepsilon_1,\varepsilon_2,\varepsilon_3)$ are mutually independent.
The average treatment effect can be identified in the Neyman-Rubin framework by adjusting the observed outcome $Y$ for covariates $C_1$ and $C_2. $

In the graphical framework, though, cyclic relationships are prohibited in DAGs, as the name itself implies.  This restricts their applicability in certain domains.  For example, in a case study on signaling networks in cells, \citet{sachs2005causal} stated: ``One of the caveats in the use of Bayesian networks for the elucidation of signaling pathways is that they are restricted to be acyclic, whereas signaling pathways are known to be rich in feedback loops".

At the same time, when cyclic relationships between variables are irrelevant for the identification of the causal effect, the variables with cyclic relationships can be represented as one node in a DAG.  Here, $C_1$ and $C_2$ can be represented by one node $\bf{C}$ (see \Cref{fig: example_cycle} and \Cref{fig: example_cluster}). In such a graph with multivariate $\bf{C},$ the traditional back-door criterion holds, and the average causal effect of $A$ on $Y$ is identified by adjusting $Y$ for $C_1$ and $C_2.$ Clustering may not help if the cycle is relevant for the identification of the causal effect. 

Moreover, if cycles arise from time-dependent relationships, they can frequently be circumvented by using distinct variables for each time point. Also,  \citet{rothenhausler2015backshift} and \citet{hyttinen2013discovering} developed the methodology to learn linear causal cyclic models in the presence of latent variables in graphical models. More recently, \citet{forre2020causal} introduced input/output structural causal models within the graphical framework that allow cycles.

\subsection*{Example 3. Deterministic relationships} 
In certain scenarios, one variable can be fully determined by other variables. For instance, in survival analyses, the observed survival is a deterministic function of censoring and the true survival time. \citet{arnold2020causal} and \citet{berrie2025depicting} discuss challenges in identification and estimation of causal effects when derived variables are involved, for instance, when birth weight determines the classification of macrosomia or when working with compositional data, where parts sum to a whole.

Consider, for example,  an i.i.d. sample from the data-generating mechanism:
\begin{align} \label{eq:model_deterministic} 
    \begin{cases}
        (C_2, C_3,\varepsilon_0,\varepsilon_1)' \sim \mathcal{N} \left(0, I_4 \right)\\
        C_1 = C_2 + C_3\\
        C_4 = C_2 - C_3\\
        A \sim Bern \left(\frac{\exp(C_1)}{1 + \exp(C_1)}\right)\\
        Y(0) = 4 + C_4 + \varepsilon_0\\
        Y(1) = 2 + C_4 + \varepsilon_1\\
        Y = AY(1) + (1-A)Y(0).\\
    \end{cases}
\end{align}
Here, as in Example 1, the consistency, positivity, and SUTVA assumptions are fulfilled. The potential outcomes are independent of the treatment  $(Y(0), Y(1)) \indep A|C_2, C_3 $ (also given $C_1$ or given $C_4$).
Therefore, the average treatment effect can be identified in the Neyman-Rubin framework through a regression of the observed outcome $Y$ on treatment $A$ and pre-treatment covariates $C_2, C_3.$

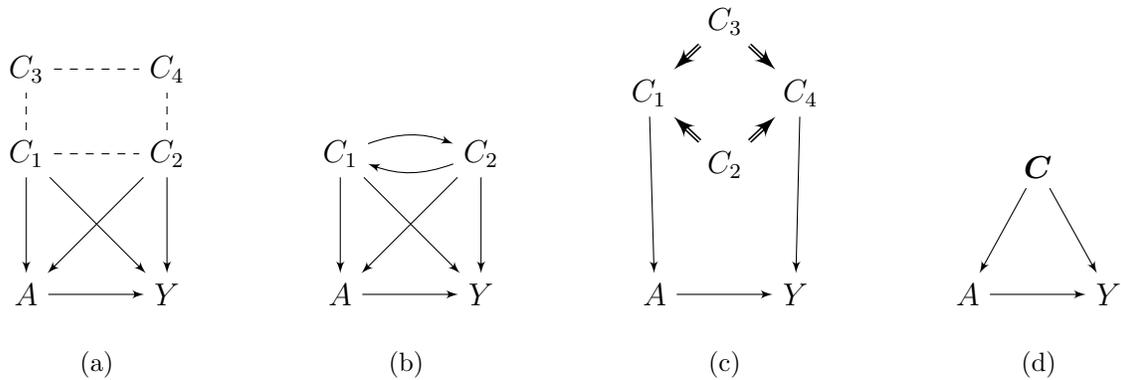
\begin{figure}
\vspace{-1cm}
\begin{center}
\subfloat[\label{fig: example_undirected}]{
\begin{minipage}[b]{0.24\textwidth}
    \centering
    \begin{tikzpicture}
    \tikzset{line width=0.6pt, outer sep=0pt,
    node distance=0.5cm}
        \node(A)       [] {$A$};
        \node(empty)   [right=of A] {};
        \node(empty_above_empty)   [above=of empty] {};
        \node(Y)       [right=of empty]{$Y$};
        \node(empty_above_A) [above=of A]{};
        \node(empty_above_Y) [above=of Y]{};
        \node(C1) [above=of empty_above_A]{$C_1$};
        \node(C2) [above=of empty_above_Y]{$C_2$};
        \node(C3) [above=of C1]{$C_3$};
        \node(C4) [above=of C2]{$C_4$};
        \draw[edge] (A) -- (Y);
        \draw[edge] (C1) -- (A);
        \draw[edge] (C1) -- (Y);
        \draw[edge] (C2) -- (A);
        \draw[edge] (C2) -- (Y); 
        \draw[dashed]  (C1) -- (C2);
        \draw[dashed]  (C1) -- (C3);
        \draw[dashed]  (C2) -- (C4);
        \draw[dashed]  (C3) -- (C4);
    \end{tikzpicture}
    \end{minipage}
}
\subfloat[ \label{fig: example_cycle}]{
\begin{minipage}[b]{0.24\textwidth}
    \centering
    \begin{tikzpicture}
    \tikzset{line width=0.6pt, outer sep=0pt,
    ell/.style={draw,fill=white, inner sep=2pt,
    line width=0.6pt},
    swig vsplit={gap=2pt},
    node distance=0.5cm}
        \node(A)       [] {$A$};
        \node(empty)   [right=of A] {};
        \node(Y)       [right=of empty]{$Y$};
        \node(empty_a) [above=of A] {};
        \node(C1)      [above=of empty_a] {$C_1$};
        \node(empty_y) [above=of Y] {};
        \node(C2)      [above=of empty_y] {$C_2$};   
        \node(empty2) [above= of C1]{};
        \node(empty3) [above= of empty2]{};
        \node(empty2) [above= of C1]{};
        \node(empty3) [above= of empty2]{};
        \draw[edge] (C1) -- (A);
        \draw[edge] (C1) -- (Y);
        \draw[edge] (C2) -- (A);
        \draw[edge] (C2) -- (Y); 
        \draw[edge] (C1) to[out=20,in=160] (C2);
        \draw[edge] (C2) to[out=200,in=340]  (C1);
        \draw[edge] (A) -- (Y);
    \end{tikzpicture}
\end{minipage}
}
\subfloat[\label{fig: example_deterministic}]{
\begin{minipage}[b]{0.24\textwidth}
    \centering
    \begin{tikzpicture}
    \tikzset{line width=0.6pt, outer sep=0pt,
    node distance=0.5cm}
        \node(A)       [] {$A$};
        \node(empty)   [right=of A] {};
        \node(empty_above_empty)   [above=of empty] {};
        \node(Y)       [right=of empty]{$Y$};
        \node(C2)      [above=of empty_above_empty] {$C_2$};
        \node(empty_above_c2) [above=of C2] {};
        \node(C3)      [above=of empty_above_c2] {$C_3$};
        \node(C1)      [left=of empty_above_c2] {$C_1$};
        \node(C4)      [right=of empty_above_c2] {$C_4$};
        \draw[edge] (A) -- (Y);
        \draw[edge] (C1) -- (A);
        \draw[edge] (C4) -- (Y); 
        \draw[edge, double]  (C3) -- (C1);
        \draw[edge, double]  (C3) -- (C4);
        \draw[edge, double]  (C2) -- (C4);
        \draw[edge, double]  (C2) -- (C1);
    \end{tikzpicture}  
\end{minipage}
}
\subfloat[\label{fig: example_cluster}]{
\begin{minipage}[b]{0.24\textwidth}
    \centering
    \begin{tikzpicture}
    \tikzset{line width=0.6pt, outer sep=0pt,
    node distance=0.5cm}
        \node(A)       [] {$A$};
        \node(empty)   [right=of A] {};
        \node(Y)       [right=of empty]{$Y$};
        \node(empty1) [above= of empty]{};
        \node(C) [above= of empty1]{$\bm{C}$};
        \node(empty2) [above= of C]{};
        \node(empty3) [above= of empty2]{};
        \draw[edge] (A) -- (Y);
        \draw[edge] (C) -- (A);
        \draw[edge] (C) -- (Y);
    \end{tikzpicture}
\end{minipage}    
}
\end{center}
\caption{Graphical representations for Examples 1-3. (a): a model with undirected relationships between $C_1, \ldots, C_4$, (b): a model with a cyclic relationship between $C_1$ and $C_2$; (c): a model with a deterministic relationship; (d) A DAG that describes the models in Examples 1-3, where variables $C_1-C_4$ are represented by one node $\bm{C}.$   In each graph, vertex $A$ represents the treatment, while $Y(a)$ represents the potential outcome under treatment level $a$. Directed edges symbolize causal effects, undirected dashed edges in (a) denote non-causal probabilistic, and double-lined edges in (c) indicate deterministic relationships. }
\label{fig:model}
\end{figure}

In both frameworks, it is important to carefully choose the target estimand and what variables, parents or children, to condition on \citep{berrie2025depicting}. 
However, additional challenges due to deterministic relationships can arise in the graphical framework when one wants to construct a DAG from observed data. 
That is because standard tools for causal discovery, such as the PC algorithm \citep{spirtes1991algorithm}, draw a DAG based on testing conditional independencies between the variables, assuming that the distribution is faithful to the graph, that is, if variables are (conditionally) independent, they are d-separated on the corresponding DAG. 
In the example, $C_1$ and $C_4$ are statistically independent (since $\text{cov}(C_1, C_4) = \text{var}(C_2) - \text{var}(C_3) = 0$ and they are jointly normally distributed), but they are not d-separated in the graph that represents the data-generating mechanism (due to confounding paths through $C_2$ and $C_3$, see \Cref{fig: example_deterministic}).  
This example illustrates
 \citet[p. 79]{hernan2020causal} statement, ``Finally, faithfulness may be violated when there exist deterministic relations between variables on the graph. 
Specifically, when two variables are linked by paths that include deterministic arrows, then the two variables are independent if all paths between them are blocked, but might also be independent even if some paths are open". 
Therefore, data-generating mechanisms with deterministic relationships between variables require much care within the graphical framework as they might violate the key assumption of faithfulness.


To address the issues caused by deterministic relationships within the graphical framework, the following methods have been developed.
Deterministic node reduction algorithm \citep{shachter1988probabilistic} enables removing deterministic variables from some graphs without losing the information (for example, in the considered example, one can remove $C_1$ and $C_4$). The D-separation criterion \citep{geiger1990identifying}, an enhanced version of the traditional d-separation,  was developed to be applied to the network with deterministic functions. \citet{mateescu2008mixed} defined mixed networks that allow for deterministic relations. \citet{chen2024identifying} considered functional projections to exclude some variables with deterministic relationships in a graph without changing the identification of causal effects. 

\section{Examples where the graphical framework excels} \label{section:dags_easier_that_po}

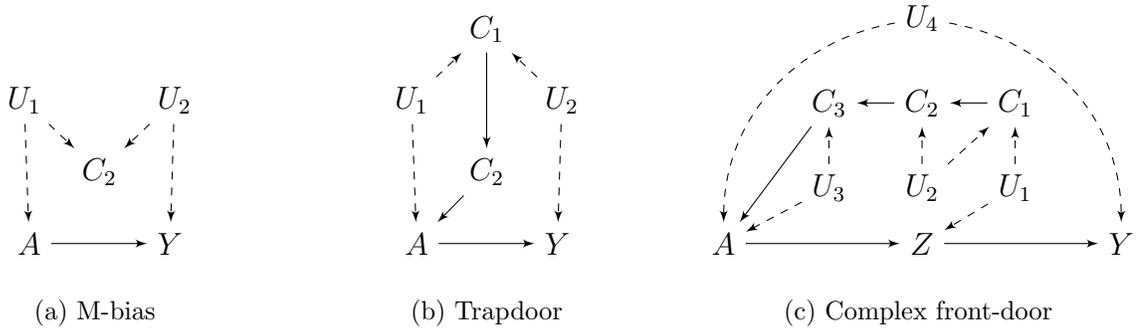
\begin{figure}
\vspace{-1cm}
\begin{center}
\subfloat[M-bias \label{fig:mbias}]{
\begin{minipage}[b]{0.3\textwidth}
    \centering
    \begin{tikzpicture}
    \tikzset{line width=0.6pt, outer sep=0pt,
    node distance=0.5cm}
        \node(A)       [] {$A$};
        \node(empty)   [right=of A] {};
        \node(Y)       [right=of empty]{$Y$};
        \node(C)      [above=of empty] {$C_2$};
        \node(empty_c) [above=of C] {};
        \node(empty_c1) [above=of empty_c]{};
        \node(U1) [left=of empty_c]{$U_1$};
        \node(U2) [right=of empty_c]{$U_2$};
        \draw[edge] (A) -- (Y);
        \draw[edge, dashed] (U1) -- (A);
        \draw[edge, dashed] (U2) -- (Y); 
        \draw[edge, dashed]  (U1) -- (C);
        \draw[edge, dashed]  (U2) -- (C);
    \end{tikzpicture}
\end{minipage}
}
\subfloat[Trapdoor\label{fig:trapdoor}]{
\begin{minipage}[b]{0.3\textwidth}
    \centering
    \begin{tikzpicture}
    \tikzset{line width=0.6pt, outer sep=0pt,
    node distance=0.5cm}
        \node(A)       [] {$A$};
        \node(empty)   [right=of A] {};
        \node(Y)       [right=of empty]{$Y$};
        \node(C2)      [above=of empty] {$C_2$};
        \node(empty_c) [above=of C2] {};
        \node[name=C1,  above=of empty_c]{$C_1$};
        \node(U1) [left=of empty_c]{$U_1$};
        \node(U2) [right=of empty_c]{$U_2$};
        \draw[edge] (C2) -- (A);
        \draw[edge]  (C1) -- (C2);
        \draw[edge] (A) -- (Y);
        \draw[edge, dashed]  (U1) -- (C1);
        \draw[edge, dashed]  (U1) -- (A);
        \draw[edge, dashed]  (U2) -- (C1);
        \draw[edge, dashed]  (U2) -- (Y);
    \end{tikzpicture}    
\end{minipage}
}
\subfloat[Complex front-door     \label{fig:complexfrontdoor}]{
\begin{minipage}[b]{0.38\textwidth}
    \centering
    \begin{tikzpicture}
        \tikzset{line width=0.6pt, outer sep=0pt,
    node distance=0.5cm}
        \node(A)       [] {$A$};
        \node(emptya1)   [right=of A] {};
        \node(emptya2)   [right=of emptya1] {};
        \node(Z)   [right=of emptya2] {$Z$};
        \node(emptyz1)   [right=of Z] {};
        \node(emptyz2)   [right=of emptyz1] {};
        \node(emptyz3)   [above=of Z] {};
        \node(Y)       [right=of emptyz2]{$Y$};
        \node(C2)  [above=of emptyz3] {$C_2$};
        \node(C1)  [right=of C2] {$C_1$};
        \node(C3)  [left=of C2] {$C_3$};
        \node(U1)  [below=of C1] {$U_1$};
        \node(U2)  [below=of C2] {$U_2$};
        \node(U3)  [below=of C3] {$U_3$};
        \node(U4)  [above=of C2] {$U_4$};
        \draw[edge] (C3) -- (A);
        \draw[edge] (A) -- (Z);
        \draw[edge] (Z) -- (Y);
        \draw[edge] (C1) -- (C2); 
        \draw[edge] (C2) -- (C3);
        \draw[edge, dashed] (U1) to (C1);
        \draw[edge, dashed] (U1) to (Z);
        \draw[edge, dashed] (U2) to (C2);
        \draw[edge, dashed] (U2) to (C1);
        \draw[edge, dashed] (U3) to (C3);
        \draw[edge, dashed] (U3) to (A);
        \draw[edge, dashed] (U4) to[out = 190, in = 90] (A);
        \draw[edge, dashed] (U4) to[out = -10, in = 90] (Y);
    \end{tikzpicture}

\end{minipage}
}
\end{center}
\caption{  
Graphs that can be handled with the graphical causal framework but might be challenging for the Neyman-Rubin framework. Variables $U_1,\ldots, U_4$ are unobserved, dashed arrows represent unobserved confounding.} 
\label{fig:challenginggraphs}
\end{figure}

Figure~\ref{fig:challenginggraphs} illustrates data-generating mechanisms that may benefit from the analyses in the graphical framework. Examples 4-6 below discuss these cases in more detail.

\subsection*{Example 4. M-bias}

Typically, to estimate the causal effect in the Neyman-Rubin framework, one would adjust for pre-treatment covariates. 
Consider a graph on \Cref{fig:mbias} where $U_1$ and $U_2$ are unobserved.  
Adjustment for covariate $C$ would unblock the path from $A$ to $Y$  \citep{greenland1999causal, sjolander2009propensity}  and, therefore, result in a biased estimator of the average causal effect. 
This bias is called a collider bias or ``M-bias" due to the structure of the corresponding DAG \citep{cole2010illustrating, griffith2020collider, zetterstrom2022selection}. 
Here, specifying a DAG as in \Cref{fig:mbias} allows us to see that one
should not condition on $C$ \citep{greenland1999causal, sjolander2009propensity}. See also Section 2.1.1.2 in \citet{markus2021causal} for a discussion of how each framework approaches the assumption of uncorrelated disturbances.

Studies based on numerical considerations and simulations \citep{ding2015adjust,liu2012implications,flanders2019limits} suggest that M-bias might not be a serious problem in practice, unless strong dependencies and non-linear relations are present.

As an extreme situation, consider the following structure with binary variables and the DAG of \Cref{fig:mbias}. Let $U_1$ and $U_2$ be unobserved biased coin tosses, i.e., $P(U_1=1)=0.6$ and $P(U_2=1)=0.4$, and define the binary variables as follows:
\begin{align*}
\begin{cases}
    & C  = (U_1 + U_2) \textrm{ mod } 2, \\ 
    & P(A=1 \vert U_1 = 1) = 0.9, \\
    & P(A=1 \vert U_1 = 0) = 0.1, \\
    & P(Y(1)=1 \vert U_2 = 1, A=a) = 0.1, \\
    & P(Y(1)=1 \vert U_2 = 0, A=a) = 0.9,\\
    & P(Y(0)=1 \vert U_2 = 1, A=a) = 0.9, \\
    & P(Y(0)=1 \vert U_2 = 0, A=a) = 0.1.
\end{cases}
\end{align*}

The average treatment effect is $E(Y(1)- Y(0)) = - 0.16.$ However, adjusting for $C$  leads to the result:
\begin{align*}
& E(Y(1)- Y(0)) = P(Y=1 \vert C=1,A=1)P(C=1) + P(Y=1 \vert C=0,A=1)P(C=0) \\
& - \left( P(Y=0 \vert C=1,A=1)P(C=1) + P(Y=0 \vert C=0,A=1)P(C=0)\right) = -0.06,
\end{align*}
which is a major divergence from the correct value.

M-bias can also appear in studies when selection criteria are applied to define the study population \citep{cole2010illustrating, griffith2020collider, zetterstrom2022selection}.
For example, \citet{zetterstrom2022selection} investigated the effect of children's very preterm birth on type 1 diabetes mellitus. 
In their analysis, the selection criteria for the study population included mothers being non-diabetic ($C$ on \Cref{fig:mbias}). 
Both the treatment (preterm delivery, $A$ on \Cref{fig:mbias}) and the selection variable (non-diabetic mother) might be influenced by the mother’s unobserved socioeconomic status ($U_1$ on \Cref{fig:mbias}). 
Also, the outcome (the child’s type 1 diabetes status, $Y$ on \Cref{fig:mbias}) and the selection variable (non-diabetic mother, $C$ on \Cref{fig:mbias}) can be influenced by mother's unobserved genes ($U_2$ on \Cref{fig:mbias}). 
\citet{zetterstrom2022selection} showed that the treatment effect for non-diabetic mothers differs from the effect in the total population of both diabetic and non-diabetic mothers.
This is because considering a subpopulation of non-diabetic mothers can be seen as conditioning on $C$ in the estimation of the causal effect, which opens backdoor paths between the treatment and the outcome.


\subsection*{Example 5. Trapdoor variables} 
\Cref{fig:trapdoor}, similar to Figure 2c in \citet{helske2021estimation}, also provides a DAG where the causal effect of $A$ on $Y$ cannot be identified by a simple adjustment in the Neyman-Rubin causal framework. 
The path $A \leftarrow C_2 \leftarrow C_1 \leftrightarrow Y$ would require conditioning on $C_1$ or $C_2$ (or both), but this would open the collider path $A \leftrightarrow  C_1 \leftrightarrow Y.$  Nevertheless, the graphical framework can help with the identification of the causal effect \citep{helske2021estimation} using the expression:
\begin{equation*} \label{eq:trapdoor}
  P(Y(a)=y) =  \frac{\sum_{c_1}P(y|c_1,c_2,a)P(a|c_1,c_2)P(c_1)}{\sum_{c_1}P(a|c_1,c_2)P(c_1)}. 
\end{equation*}
Fixing the value of the trapdoor variable $C_2$ is required for identification. The small sample properties of different choices of $C_2$ are studied by 
\citet{helske2021estimation} also offers an example of a trapdoor structure in applied research by studying the effect of education on income in Finland. In their model, the primary school grade point average is a trapdoor variable.

\subsection*{Example 6. Complex front-door}

\begin{figure}
\vspace{-1cm}
\begin{center}
\subfloat[\centering Back-door adjustment ]{
    \begin{minipage}[b]{0.3\textwidth}
    \centering
    \begin{tikzpicture}
    \tikzset{line width=0.6pt, outer sep=0pt,
    node distance=0.5cm}
        \node(A)       [] {$A$};
        \node(empty)   [right=of A] {};
        \node(Y)       [right=of empty]{$Y$};
        \node(C) [above= of empty]{$C$};
        \draw[edge] (A) -- (Y);
        \draw[edge] (C) -- (A);
        \draw[edge] (C) -- (Y);
    \end{tikzpicture}
    \end{minipage}
}
\subfloat[\centering Front-door adjustment    \label{fig:frontdoor}]{
    \begin{minipage}[b]{0.3\textwidth}
    \centering
    \begin{tikzpicture}
    \tikzset{line width=0.6pt, outer sep=0pt,
    node distance=0.5cm}
        \node(A) [] {$A$};
        \node(Z) [right=of A] {$Z$};
        \node(Y) [right=of Z]{$Y$};
        \node(empty_z) [above= of Z]{};
        \node(U) [above= of empty_z]{$U$};
        \draw[edge] (A) -- (Z);
        \draw[edge] (Z) -- (Y);
        \draw[edge, dashed] (U) -- (A);
        \draw[edge, dashed] (U) -- (Y);
    \end{tikzpicture}
    \end{minipage}
    }
\subfloat[\centering Front-door with a pre-mediator confounder \label{fig:frontdoorconfounding}]{
    \begin{minipage}[b]{0.3\textwidth}
    \centering
    \begin{tikzpicture}
    \tikzset{line width=0.6pt, outer sep=0pt,
    node distance=0.5cm}
        \node(A) [] {$A$};
        \node(Z) [right=of A] {$Z$};
        \node(Y) [right=of Z]{$Y$};
        \node(empty_z) [above= of Z]{};
        \node(U) [above= of empty_z]{$U$};
        \node(C) [above=of Z] {$C$};
        \draw[edge] (A) -- (Z);
        \draw[edge] (Z) -- (Y);
        \draw[edge, dashed] (U) -- (A);
        \draw[edge, dashed] (U) -- (Y);
        \draw[edge] (C) -- (A);
        \draw[edge] (C) -- (Z);
    \end{tikzpicture} 
    \end{minipage}
}
\end{center}
\caption{ Data-generating mechanisms that have readily available tools to estimate causal effects in the Neyman-Rubin causal framework. Dashed arrows represent unobserved confounding.}
\label{fig:N-R available}
\end{figure}
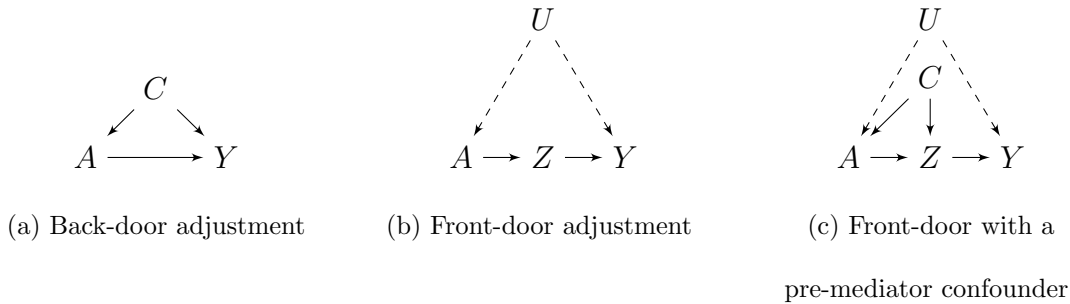

The Neyman-Rubin approach has readily available tools to estimate causal effects by back-door, front-door, and generalized front-door (two-door) adjustments for the systems represented, for example, in \autoref{fig:N-R available} (\citealp{rosenbaum1983central,fulcher2020robust,gorbach2023contrasting}). When the number of variables is moderate or large and the causal relations are complicated, a long list of assumptions is needed to specify the model in the Neyman-Rubin causal framework.  DAGs and the graphical approach can convey the same information more concisely.
\Cref{fig:frontdoor} and \Cref{fig:frontdoorconfounding} represent graphs in which the average causal effect is identified in the Neyman-Rubin framework via front-door and generalized front-door adjustments \citep{fulcher2020robust, gorbach2023contrasting}. The use of the Neyman-Rubin framework becomes tedious when the number of variables increases. 

For example, the observed confounders $C_1$, $C_2,$ and $C_3$ on Figure~\ref{fig:complexfrontdoor}  form a nested front-door structure that confounds the effect of $A$ on $Z$. The causal parameter $EY(a)$ is still identifiable, although through a complicated formula for $P(Y(a)=y)$ that, after automated simplification by \texttt{causaleffect} R-package \citep{tikka2017simplifying} that uses the graphical framework,  can be presented as follows: 

\begin{equation} \label{eq:complexfrontdoor}
    P(Y(a)=y) = \frac{1}{g_1(a,c_2)} \sum_{z, c_3}\left\{ g_2(a,z,c_2) \sum_{a} \left[ \frac{g_5(y,a,z,c_2,c_3) g_4(a,c_2,c_3)}{g_3(a,z,c_2,c_3) } \right] \right\},
\end{equation}
where
\begin{align*}
g_1(a,c_2) & = \sum_{c_1}P(a|c_1,c_2)P(c_1), \\
g_2(a,z,c_2) & = \sum_{c_1}P(z|c_1,c_2,a)P(a|c_1,c_2)P(c_1), \\
g_3(a,z,c_2,c_3) & =\sum_{c_1}P(z|c_1,c_2,c_3,a)P(a|c_1,c_2,c_3)P(c_3|c_1,c_2)P(c_1),\\
g_4(a,c_2,c_3) &= \sum_{c_1}P(a|c_1,c_2,c_3)P(c_3|c_1,c_2)P(c_1),\\
g_5(y,a,z,c_2,c_3) &=  \sum_{c_1}P(y|c_1,c_2,c_3,a,z)P(z|c_1,c_2,c_3,a)P(a|c_1,c_2,c_3)P(c_3|c_1,c_2)P(c_1).
 \end{align*}
 Here $C_2$ is a trapdoor variable that is necessary for the identification but can be chosen freely in \Cref{eq:complexfrontdoor}.
Although this formula could in principle be derived using the Neyman-Rubin framework, there are practical obstacles. One should be able to list all the conditional independence relations that are needed without the help of the graph in Figure~\ref{fig:complexfrontdoor} and then apply many steps of probability calculus to derive  \Cref{eq:complexfrontdoor}. Moreover, if the assumptions are then slightly changed, the whole process need to be repeated.

\section{Conclusion} \label{sec:conclusion}
This paper aimed to provide simple examples which highlight challenges faced by the Neyman-Rubin and graphical causal frameworks for causal reasoning and to illustrate how the two frameworks can  complement each other in the identification and estimation of causal effects. 
Examples 1--3 consist of data-generating mechanisms that include undirected, cyclic, and deterministic relationships that are not straightforwardly tractable using causal DAGs. However, simple covariate adjustment in the Neyman-Rubin framework can be used to identify the average causal effect. Conversely, Examples 4--6 illustrate data-generating mechanisms where identification in the Neyman-Rubin framework is not straightforward to study. For these examples, the graphical framework not only illustrates the underlying systems but also provides tools for the identification of the causal effect of interest.  

This paper does not attempt to provide an exhaustive comparison of the Neyman-Rubin and graphical frameworks. For those interested in exploring this relationship further, apart from the aforementioned papers, there is extensive literature available, including works by \citet{shpitser2016causal} and \citet{malinsky2019potential}.

Nevertheless, the examples presented here may offer researchers and a wider audience valuable insights into how these two widely used causal inference frameworks can complement each other and enhance causal inference research.

\section*{Funding}
This work was supported by the Marianne and Marcus Wallenberg Foundation  under Grant number MMW 2021.00207, the Swedish Research Council under grant number 2016-00703, and by the Research Council of Finland under grant number 368935.

\section*{Disclosure statement}
The authors report there are no competing interests to declare.

\bibliographystyle{apalike}

\end{document}